# Temperature-driven confinements of surface electrons and adatoms in a weakly interacting 2D organic porous network

Lu Lyu[1,*], Jin Xiao[2], Zakaria M. Abd El-Fattah[3], Tobias Eul[1], Mostafa Ashoush[3], Jun He[2], Wei Yao[1], Ignacio Piquero-Zulaica[4], Sina Mousavion[1], Benito Arnoldi[1], Sebastian Becker[1], Johannes V. Barth[4], Martin Aeschlimann[1], Benjamin Stadtmüller[1,5,*]

[1]Department of Physics and Research Center OPTIMAS, Rheinland-Pfälzische Technische Universität Kaiserslautern-Landau, Erwin-Schrödinger-Straße 46, 67663 Kaiserslautern, Germany.
[2]School of Science, Hunan University of Technology, Zhuzhou 412007, China.
[3]Physics Department, Faculty of Science, Al-Azhar University, Nasr City, E-11884 Cairo, Egypt.
[4]Physics Department E20, Technical University of Munich, 85748 Garching, Germany.
[5]Institute of Physics, Johannes Gutenberg University Mainz, Staudingerweg 7, 55128 Mainz, Germany.

[*]Email: llyu@rhrk.uni-kl.de
[*]Email: bstadtmueller@physik.uni-kl.de

**Abstract**

Two-dimensional organic porous networks (2DOPNs) have opened new vistas for tailoring the physicochemical characteristics of metallic surfaces. These typically chemically bound nanoporous structures act as periodical quantum wells leading to the 2D confinements of surface electron gases, adatoms and molecular guests. Here we propose a new type of porous network with weakly interacting 2,4,6-triphenyl-1,3,5-triazine (TPT) molecules on a Cu(111) surface, in which a temperature-driven ($T$-driven) phase transition can reversibly alter the supramolecular structures from a close-packed (CP-TPT) phase to a porous-network (PN-TPT) phase. Crucially, only the low-temperature PN-TPT exhibits subnano-scale cavities that can confine the surface state electrons and metal adatoms. The confined surface electrons undergo a significant electronic band renormalization. To activate the spin degree of freedom, the $T$-driven PN-TPT structure can additionally trap Co atoms within the cavities, forming highly ordered quantum dots. Our theoretical simulation reveals a complex spin carrier transfer from the confined Co cluster to the neighbouring TPT molecules via the underlying substrate. Our results demonstrate that weakly interacting 2DOPN offers a unique quantum switch capable of steering and controlling electrons and spin at surfaces via tailored quantum confinements.

**Introduction**

Two-dimensional organic porous networks (2DOPNs) have been identified as versatile nanoscale platforms for exotic tessellation patterns[1], catalytic reactions[2], magnetic nanostructures[3,4], and quantum confinement at surfaces[5]. Beyond the pioneering artificial corral structures formed by single atom/molecule manipulation[6,7], surface molecular self-assembly can trigger the formation of extended and well-defined 2DOPNs[8]. The protocol of self-assembled 2DOPNs relies on selecting an appropriate organic species with defined end-groups that facilitate the bonding with other molecules. The coordinated bonds can be mediated by robust covalent[9], metal-ligand[10] and hydrogen/halogen-bonded[11,12] interactions. The resulting 2DOPNs exhibit periodic nanocavities that can reshape the metal surface potential landscape[13]. In practice, an organic molecule can create a potential barrier on a surface to partially block the penetration of surface electrons and atoms[14], as reported for the adatom-induced interference to the surface state (SS) electrons[15]. Therefore, 2DOPN cavities form a periodic array of quantum wells by organic building blocks, and some experiments of scanning tunneling microscopy (STM) and angle-resolved photoemission (ARPES) have confirmed the existence of confined states for surface electrons and adsorbates[5]. In addition, organic porous confinements can further be modulated by the structural environments, such as blocks barrier width[12], porous geometry[16] and adsorbate occupancy[17].

As mentioned above, the 2DOPNs are always constructed from the directional and robust bondings. In this article, however, temperature-driven quantum confinement is proposed in a weakly interacting organic system, i.e., 2,4,6-triphenyl-1,3,5-triazine (TPT) molecules on Cu(111) surface. Our previous study[18] uncovered that the external temperature ($T$) acts as a thermal switch to steer the TPT architectures between a close-packed phase at room temperature ($RT$) and a porous-network phase at low temperature (< 170 K). In the $T$-driven porous network, we demonstrate the tunability of the quantum confinements to the SS electrons and magnetic Co adatoms. Furthermore, the confined Co atoms can activate the spin degree of freedom within the porous-network. DFT calculations shed light on the subtle spin transfer process. The spin-dependent charges transfer from the porous Co cluster to the neighbouring TPT molecules via the Cu substrate.

# Results

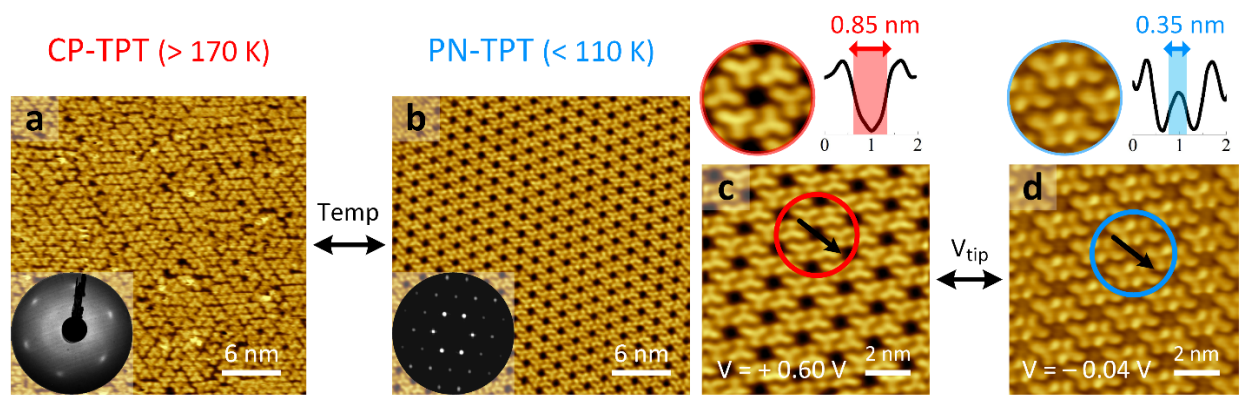

**Fig. 1 | Quantum confinement of surface electrons in TPT porous network on Cu(111).** STM topographies at tip voltage ($V_{tip}$) of + 0.60 V show a reversible phase transition between (**a**) close-packing (CP-TPT, 170 < T < 297 K ) and (**b**) porous-network (PN-TPT, T < 110 K). The corner insets of **a,b** are the LEED (energy at 12 eV) and the fast Fourier transform (25 × 25 nm$^{-2}$), respectively. Highly-resolved structures of the PN-TPT are achieved at the $V_{tip}$ of + 0.60 (**c**) and − 0.04 V (**d**). On the top of **c,b**, the magnified porous units (in the red and blue circle) and the line profiles crossing the cavity (along the black arrow) are located at the left and the right. The line profiles indicate an empty and domelike porous structure with the FWHM of 0.85 nm and 0.35 nm, respectively.

**Quantum confinement of surface electrons in TPT porous network.** In Fig. 1a,b, A *T*-driven phase transition is observed in the 0.8 monolayer (ML) TPT film on Cu(111). At the *RT*-STM (Fig. 1a), the TPT molecules form a close-packed structure (CP-TPT). The low energy electron diffraction pattern exhibits a *p*(3 × 3) hexagonal superlattice originating from the locally ordered TPT phenyl groups[18]. Cooling the sample to 106 K, the STM (Fig. 1b) and the Fast Fourier Transform (FFT) image show a long-range ordered porous network (PN-TPT). The *T*-driven structures show a reversible phase transition, and the mixed phases (Fig. SI1) are observed in some intermediate temperatures (110 < *T* < 170 K).

In the small-scale STM image (Fig. 1c), it clearly shows that the "flower-like" porous units constitute a periodic PN-TPT network, and each cavity contains six TPT blocks. At the tip bias ($V_{tip}$) of + 0.60 V, the line profile extracted along the black arrow shows an empty cavity with a size of 0.85 nm (in subnanoscale). When the $V_{tip}$ is changed to − 0.04 V, a domelike state (Fig. 1d) emerges at every porous center, where a protruding peak (size of 0.35 nm) is shown in the line profile. From the previous study by Lobo-Checa et al.[19], this in-cavity feature is the characteristic signature of the confined SS electrons. Therefore, the porous state appears near the Fermi energy ($V_{tip}$ close to zero), and a series of the $V_{tip}$-dependent porous states are shown in Fig. SI2.

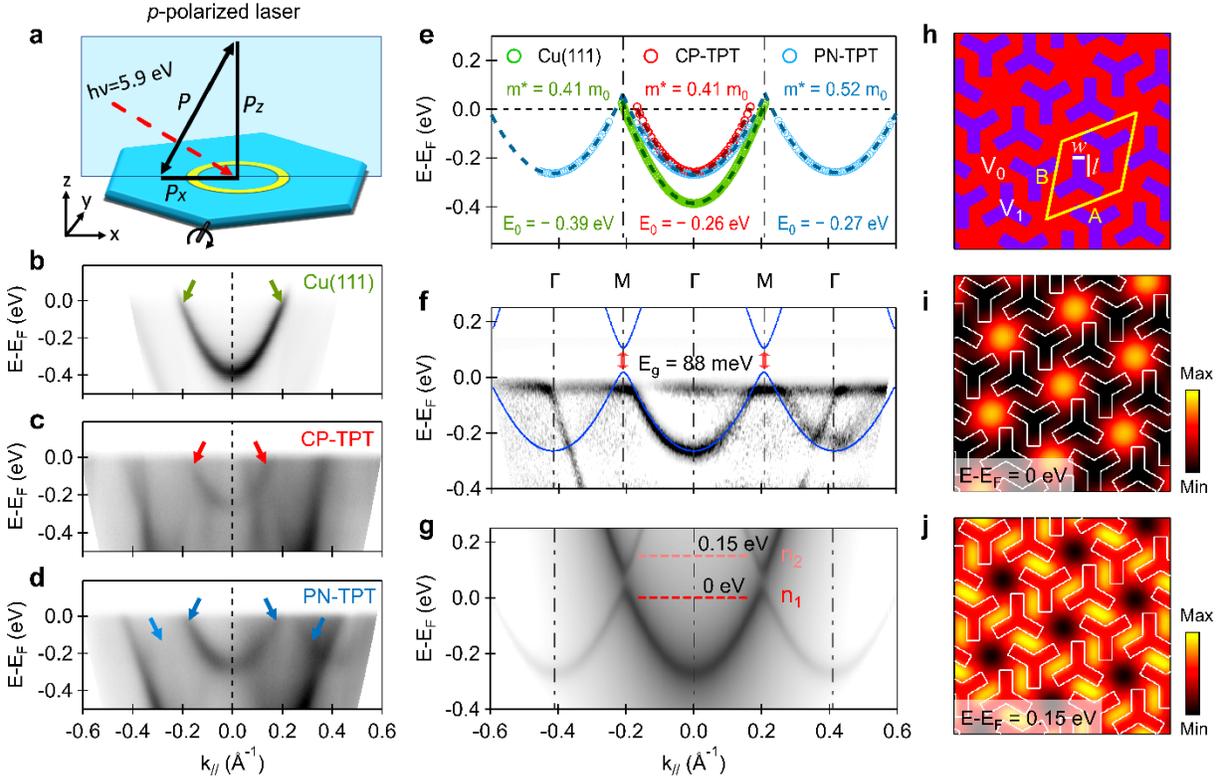

**Fig. 2 | *T*-driven surface states confinement in TPT porous network. a**, A schematic of linearly *p*-polarized laser-ARPES setup (hν = 5.9 eV). **b-d**, ARPES band structures of Cu(111), CP-TPT (at RT), and PN-TPT (at 30 K) near the Fermi. The coloured arrows mark the surface state (SS) band structures. **e**, The coloured circles show the positions of the SS bands in **b-d**, that are extracted from the maximum intensities in the fitted energy distribution curves. The overlapped dash parabolas are from the SS dispersion relation, and the effective mass ($m^*$) and minimum energy ($E_0$) are obtained and labelled. **f**, The second derivative map of **d**, improving the visualization of the SS replicas on the left and right. **g**, The electron plane wave expansion (EPWE) simulated SS band structures in PN-TPT, and the band positions are extracted and overlapped in **f** (blue curves). **h**, PN-TPT model for the EPWE simulations. The geometry parameters, according to the STM structure, are set as |**A**| = |**B**| = 17.85 Å, molecular size $w = l/\sqrt{3}$ = 3.0 Å. The potential of the substrate (red area) and the triangular TPT (blue parts) are $V_0$ = 0 meV and $V_1$ = 454 meV, respectively. **i,j**, EPWE simulations of the SS local density of state (LDOS) maps in PN-TPT, taken at $E-E_F$ = 0 eV and 0.15 eV, respectively. The two energies are located in the two discrete eigenstates ($n_1$ and $n_2$), marked in **g**. All the $k_{//}$ in band structures are along the $\overline{\Gamma M}$ direction in PN-TPT. The scales of **h-j**, 5 × 5 nm$^2$.

**Surface state (SS) in TPT porous network.** Due to a surface symmetry broken of noble metals in the perpendicular direction[20], the associated SS exists and propagates like nearly-free electron gas on the metal surface. SS is located in the Γ-L projected bulk band gap[21], and the parabolic dispersion of the SSs follows the relation $E(k_{//}) - E_F = E_0 + \frac{\hbar^2 k_{//}^2}{2m^*}$, where $E_F$, $E_0$, $m^*$ and $k_{//}$ are the Fermi energy, band minimum, the effective band mass and parallel momentum wave-vector. In experiments, the dispersion relation can be quantified from ARPES measurements, and a p-polarized light (Fig. 2a) can effectively excite SSs electrons[22]. In Fig. 2b-d, The ARPES maps show the electronic band structures of bare Cu(111), CP-TPT, and PN-TPT close to $E_F$. By altering the laser polarizations (p- to s-, details in Fig. SI3e), SS bands can be identified, as arrow marked in Fig. 2b-d. In Fig. 2e, The positions of the SS bands are extracted from the three structures, and they are fitted with the parabolic dispersion (dashed curves) to obtain the parameters of $m^*$ and $E_0$. In the bare Cu(111), the well-known SS has the $m^* = 0.41\ m_0$ ($m_0$, free electron mass) and the $E_0 = -0.39$ eV. After the adsorption of TPT molecules at $RT$, the $m^*$ remains unchanged in the CP-TPT, but the $E_0$ shifts by 0.13 eV due to a push-back effect[23]. When the structure is cooled to PN-TPT (30 K), the SS parabola opens wider and the $m^*$ increases to 0.52 $m_0$. Considering the SS electrons are confined into each cavity in PN-TPT, as seen in the STM (Fig. 1d). The $m^*$ enhancement can be attributed to the quantum confinement of the porous network. As a result, the confinement of PN-TPT induces the $m^*$ renormalization magnitude of 27% (($m^* - m_0$) / $m_0$).

In Fig. 2d, two replicated SSs are observed to the left and right of the central SS. The replicas are clearly visible in the second derivative map (Fig. 2f), where they keep a periodicity of $k_{//} = 0.41$ Å$^{-1}$, consistent with the reciprocal lattice of the PN-TPT porous network. The SSs, along with the different Brillouin-zones directions, are shown in Fig. SI3a-d. In addition, the SS bands in PN-TPT are simulated using the semiempirical electron plane wave expansion (EPWE) method. Based on the structural model (Fig. 2h), the EPWE simulated band structures, as seen in Fig. 2g, reproduce the replicated SSs. The extracted positions (blue curve) below the $E_F$ overlap well on the APRES map in Fig. 2f. Interestingly, at the crossing $M$-point (above the $E_F$), the simulation shows a gap opening of $E_g = 88$ meV. As a result, the continuous SS band changes to the discrete states (n$_1$ and n$_2$) below E − E$_F$ = 0.2 eV, and the quantized phenomenon is from the confinement effect of SS in a porous network[24]. The simulated local density of state (LDOS) spectra (Fig. SI4) indicate that the n$_1$ and n$_2$ states are located in different porous positions. Fig. 2i and 2j show the spatial distributions of the n$_1$ state (at E − E$_F$ = 0 eV) and the n$_2$ state (at E − E$_F$ = 0.15 eV) in the LDOS maps, respectively. The domelike n$_1$ states reside in each cavity, in agreement with the confined SS electrons in the STM (Fig. 1d). The n$_2$ states are mainly located in the intermolecular gaps between the cavities. It indicates that the intermolecular gap in PN-TPT is a transmission channel for the confined SS, and that the electrons can couple with the neighbouring ones through the channel[19,25]. An intercoupling effect between porous states leads to Bloch-like dispersions, as seen in Fig. 2g.

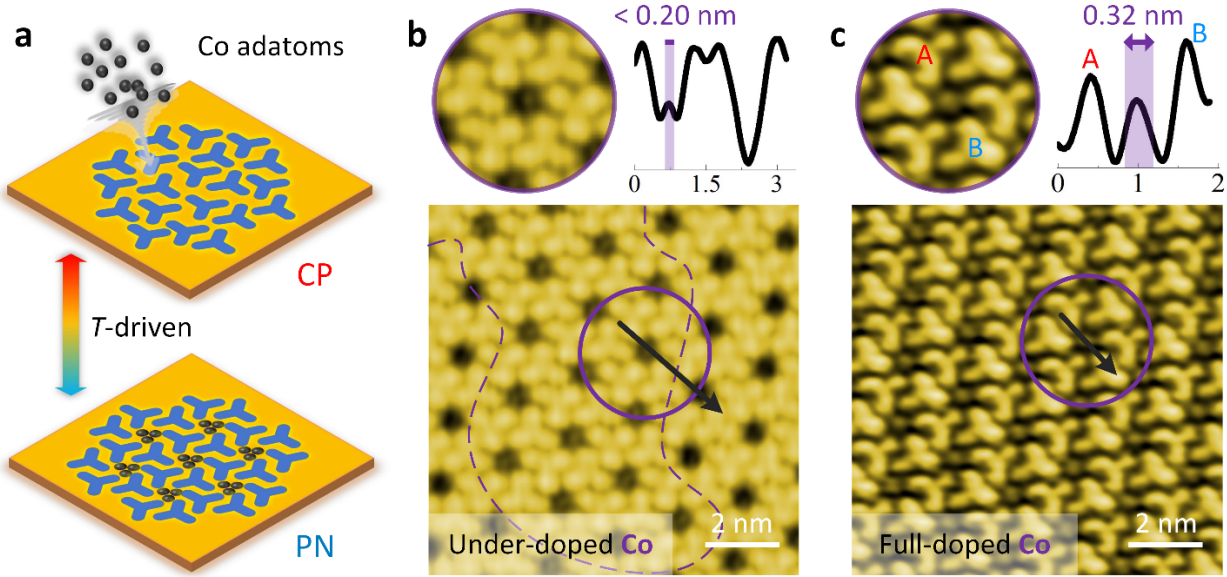

**Fig. 3 | *T*-driven adatoms confinement in TPT porous network. a**, A *T*-driven protocol for atoms confinement in the TPT porous network. Reversible process of Co adatoms between a gas phase on the CP-TPT (RT) and a condensed phase in the cavities of the PN-TPT (106 K). **b**, At low-dosing Co atoms (0.008 ML) in PN-TPT, the STM ($V_{tip}$ = + 0.30 V) shows an under-doped Co in the purple-line region. In the top-left porous unit (magnified in the purple circle), there is a tiny Co occupancy in the cavity, where the line profile (along the black arrow) shows the occupancy size < 0.20 nm. **c**, Increasing the Co dose to 0.04ML, the STM ($V_{tip}$ = + 0.30 V) shows the full-occupied Co adatoms in each cavity. In the magnified unit, the confined Co atoms form a cluster in the porous center. The line profile shows the cluster size of 0.32 nm, and two types of TPT (A and B) with different states in a full-doped unit. All STM were performed at 106 K, and the tunneling currents ($I_t$) are both 0.05 nA.

**Co adatoms in TPT porous network.** In the previous setion, The *T*-driven quantum confinement of SS electrons has been uncovered in the real and momentum spaces. In this section, a protocol of *T*-driven confinement to magnetic Co adatoms will be proposed in the TPT system. As seen in Fig. 3a, the schematic shows the controllable confinement to the Co adatoms. At *RT*, low-dose Co atoms are deposited on the CP-TPT layer. The adatoms can act as mobile gases, and the underlying CP-TPT structure is unchanged in the STM and LEED. However, cooling to the PN-TPT phase, the porous architectures enable confining the Co atoms into the cavities. Interestingly, the *T*-driven phase transition between the PN-TPT and the CP-TPT remains reversible even after depositing the small amount of metal atoms. Therefore, the Co confined process is also reversibly controlled by temperature, and the confined Co atoms can be pushed out again when the cavity is closed at an elevated temperature.

In Fig. 3b,c, the STM images show the two different doses of Co atoms in PN-TPT. At a low dose of 0.008 ML (Fig. 3b), partial cavities in the purple line region have a dot-like structure (named under-doped Co). The line profile (along with the black arrow) across two cavities shows shows a tiny peak with a width smaller than 0.20 nm in the occupied cavity and no peak within the unoccupied cavity. Increase the dose to 0.04 ML (Fig. 3c), each porous center appears an identical spot (named full-doped Co). The crossed line profile shows the full-doped Co with a larger size of 0.32 nm, indicating a Co cluster in the cavity. The signature of the full-doped Co is measured by STM (Fig. SI5) at a series of $V_{tip}$ and tunneling current ($I_t$). Interestingly, the full-doped Co leads to a porous symmetry breaking, and the surrounding TPT molecules are divided into two types, marked A (TPT$_A$) and B (TPT$_B$). The line profile indicates a lower density of state in TPT$_A$, which does not occur in the under-doped case. Further increasing the dose of 0.04 ML, the excess Co atoms destroy the low-temperature porous structure, as shown in Fig. SI6a. Besides trapping the Co atoms, the protocol of $T$-driven atom confinement also works for the nonmagnetic Cu atoms shown in Fig. SI6b-d.

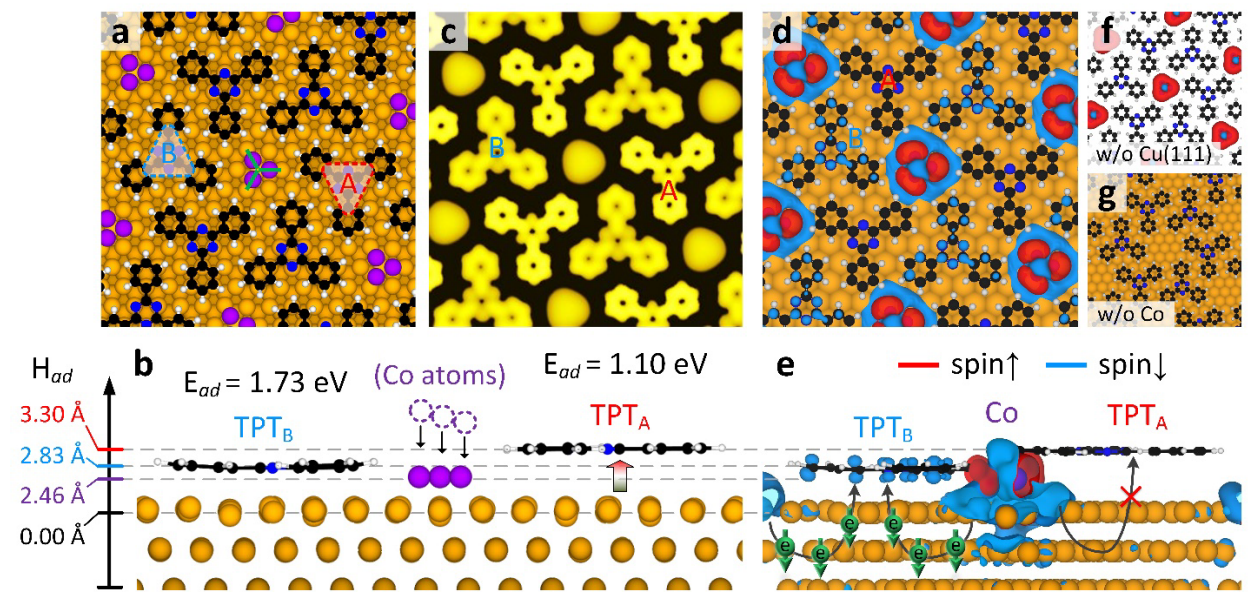

**Fig. 4 | Structure and spin simulations of full-doped Co in TPT porous network.** Top view (**a**) and side view (**b**) of the optimized structure of full-doped Co atoms in PN-TPT: three Co atoms on the porous Cu bridge sites (marked by the green bar), and two stacking types of TPT on the surface (TPT$_A$ and TPT$_B$). The adsorption heights ($H_{ad}$) and adsorption energies ($E_{ad}$) of TPT$_A$ and TPT$_B$ are labelled in **b**. The TPT$_A$ is lifted by the trapped Co atoms. **c**, A simulated STM image for **a** structure, calculated at a constant current mode with the energy of 0.6 eV below the Fermi. **d,e**, Spin density differences (SDD, majority spin (↑, red) - minority spin (↓, blue)) of full-doped Co atoms in PN-TPT in top and side view. The SDD of **f** (w/o Cu(111), removal of the Cu substrate from **d**) and **g** (w/o Co, removal of the porous Co atoms from **d**). The spin transfer path is marked in **e**: minority spin electrons from the porous Co atoms to the neighbouring TPT$_B$ via the Cu substrate. Isosurface values (e·Å$^{-3}$): 0.5 in **c**, and 0.0003 to majority spin and 0.0001 to minority spin in **d-g**. The scales of **a,c,d,f,g** are 3.6 × 3.6 nm$^2$.

**Simulations of full-doped Co in TPT porous network.** Adatom confinement in a 2DOPN is a complex process, which is dominated by temperature, substrate, molecule blocks and adjacent adatoms. We employed density functional theory (DFT) to reveal the preferred structure of Co atoms in PN-TPT. At first, the optimized structure of bare PN-TPT on Cu(111) is shown in Fig. SI7. Two types of TPT molecules are found to be adsorbed on the different positions of the Cu surface. They correspond to the TPT$_A$ and TPT$_B$ marked in Fig. 3c. Without Co doping, the TPT$_A$ has a lower adsorption energy ($E_{ad}$), but almost the same height ($H_{ad}$) as the TPT$_B$. In Fig. SI8 and SI9, different numbers of Co atoms trapped in a PN-TPT cavity are simulated for the under-doped Co and full-doped Co structures. The optimized full-doped Co structure is shown in Fig. 4a. Three Co atoms are located in the porous three symmetric bridge sites (named B$_2$ site, marked by the green bar). In Fig. 4b, the process of the Co atoms trapped in the cavity induces a lifting of the TPT$_A$ from the surface. The $H_{ad}$ of TPT$_A$ is about 0.5 Å higher than that of TPT$_B$, which indicates a weaker hybridization state between TPT$_A$ and substrate. As a result, the simulated STM image (Fig. 4c) confirms a lower state density of TPT$_A$, and the three confined Co atoms are in a unified cluster state. The calculated results agree with the STM observations in Fig. 3c.

Fig. 4d-g show the spin transfer of the full-doped Co in PN-TPT. The calculations of the charge density differences and the atom Bader charges (Fig. SI10) indicate a charge transfer from the confined Co cluster to the neighbouring TPT$_B$ molecules via the substrate. Interestingly, the spin density differences (SDD) plots in Fig. 4d,e show that the transferred charges have mainly minority spin characters, pointing to a spin polarization of the TPT$_B$ molecules. To confirm that, we remove the substrate (Fig. 4f) and porous Co atoms (Fig. 4g), respectively. The corresponding SDD show no spin transfer between the Co cluster and the TPT molecules, as well as the TPT and the Cu substrate. The pure minority spin charges from the magnetic Co atoms are injected into the neighbouring TPT$_B$ molecules mediated by the Cu substrate (paths marked in Fig. 4e). Furthermore, the strength of spin accumulation on the molecules depends on the molecule-surface coupling, and thus the uplifted TPT$_A$ molecules have negligible spin states.

**Discussion**

Our joint experimental and theoretical observations have uncovered several unique features of the weakly interacting TPT porous network: (i) the temperature of the substrate that allows a *T*-driven opening and closing of the cavities; (ii) the miniaturization of the cavity that surpasses the molecular size; (iii) high cavity density by the smaller interporous distance. These advantages enable new types of functionalities that are elusive in most chemically coordinated 2DOPNS.

Quantum confinement in PN-TPT causes a strong renormalization of the SS electrons: the domelike porous state, the shifting $E_0$ ($\Delta E_0$ = 130 meV) and the increased $m^*$ ($\Delta m^*$ = 27%). Remarkably, the EPWE simulations demonstrate a band gap opening and the continuous SS changes to the Bloch-like bands.

For adsorbates confined in a 2DOPN (a host-guest system), the guest structure is dominated by multiple interplays, such as system temperature[26], substrate interaction[27], blocks potential[28], adsorbates interaction[29] and confined surface states[30]. For adatoms in robust 2DOPNs, e.g., Xe in (3deh-DPDI)$Cu_2$[26] and Fe in (NC-$Ph_3$-CN)$_3Cu_2$)[31], the adatoms are only confined at a liquid helium temperature (< 15 K), and the gentle annealing can stimulate the occupancies hopping out of the cavity. Comparing the full-doped Co in the weakly interacting TPT porous network, the three atoms formed cluster is stable even above a liquid nitrogen temperature ($T$ < 110 K). However, adatoms in a weakly interacting porous network can alter the surrounding molecule structure. Due to the TPT blocks' repulsive potential in the cavity[18], the more weakly adsorbed $TPT_A$ molecule (lower $E_{ad}$) is easily pushed away from the surface by interacting with the Co adatoms. Meanwhile, the confined Co atoms can activate the spin degree of freedom within the PN-TPT network. DFT calculations shed light on the subtle charge and spin transfer process. The Cu substrate plays a transmission channel for the spin-dependent charges from the porous Co cluster to the neighbouring TPT molecules.

In conclusion, we propose a prototype of a controllable quantum switch on weakly interacting 2DOPNs. Based on the metal/organic interface, i.e., TPT molecules on Cu(111), the temperature manipulates the reversible structures of a phase transition between CP-TPT and PN-TPT. The $T$-driven architectures can play as a switch, "OFF" (close-packing) and "ON" (porous network), on the quantum confinements of surface state electrons and metal adatoms. The associated electronic and spintronic confinement states occur in the $T$-driven switch tuning to the "ON" (PN-TPT phase). Therefore, the weakly interacting 2DOPN offers unique functionality in quantum manipulations to tailor the surface and interface physicochemical properties.

## Methods

**Sample preparation.** All experiments were carried out in two custom-built ultra-high-vacuum systems (UHV, base pressure below $10^{-10}$ mbar) housing on Omicron VT-STM and SPECS Spectroscopy. The sample preparations were carried out under the same conditions. The monolayer TPT film was prepared by the procedure reported in our previous study[18]. Before the TPT deposition, the Cu(111) crystal was processed by several cycles of Ar-ion sputtering and subsequent annealing. The Co and Cu depositions were done using an electron beam evaporator with a low flux (6 nA). The deposition amount is calibrated by the STM to the submonolayer Co coverage on Au(111).

**Variable temperature scanning tunneling microscope (VT-STM).** The experimental setup for the VT-STM (30-500 K) experiments has been reported elsewhere[18]. In brief, all STM images were taken in constant current mode with the tunneling current ($I_t$) in the range of 70-90 pA. The bias voltage ($V_{tip}$) is applied to the STM tip, and the positive and negative values correspond to tunneling into the samples' occupied and unoccupied states, respectively. The sample temperature can be tuned by a cooling system (LakeShore 335 temperature controller). The STM images were processed with the Gwyddion software[32].

**Linearly-polarized laser photoelectron spectroscopy.** The angle-resolved photoemission spectroscopy (ARPES) measurements were carried out using a linearly polarized laser and a hemispherical analyzer (Specs PHOIBOS 150). The laser photons of 5.9 eV energy are generated from a frequency-quadrupled (4th harmonic with BBO-crystals) solid-state laser system (Ti-Saphire oscillator Griffin 10), and the laser spot size on the sample is about 92 μm. The polarization of the laser pulses can be controlled by using a λ/2-plate. Fig. 2a and Supplementary Fig. 3e show the definition of the polarizations and the experimental geometries. The angle of incidence of the laser beam on the sample is 45°, and the sample can be rotated around the manipulator polar-axis (along the y-direction) to observe a wider momentum space. The negative momentum (in $-k_{//}$ direction) refers to the sample's anticlockwise rotation, and the laser *p*-polarized vector is more components parallel to the surface (larger $p_x$ vector) like an *s*-polarization. The energy and angular resolution of the analyzer were set to better than 20 meV (at 10 eV pass energy) and 0.3° (0.01Å$^{-1}$), respectively. A sample bias voltage − 4 V was applied in the ARPES to measure the low kinetic photoelectrons, and the photoemission distortions induced by the bias were corrected[22]. An integrated Helium-cryostat allows cooling of the samples in ARPES setup to 30 K.

**EPWE simulations.** Electron Plane Wave Expansion (EPWE) method has been extensively used for 2DEGs scattering in arbitrarily shaped potential barrier blocks[15,18,33]. It is based on Green's functions for finite structures (local density of states) and expanded periodic arrays (Bloch-wave states). Depending on defining scattering geometry and periodic barrier conditions, the particle-in-a-box model can be extended to infinite 2D systems addressing the band structures. Within the EPWE code, a linear combination of plane waves is used in the solutions of the Schrödinger equation, and a satisfactory convergence was achieved with a basis set consisting of ~100 waves.

**Ab initio calculations.** Density functional theory calculations (DFT) were carried out with the VASP code[34]. The ion-core interactions were described by the projected augmented wave (PAW) method[35] using a 400 eV plane-wave cut-off, and the electron-exchange correlation was used by the generalized gradient approximation (GGA) with the Predew-Burke-Ernzerhof (PBE) function[36]. To accurately include long-range van der Waals interactions, the optB86b+vdW[37] nonlocal exchange-correlation function was applied. In the calculated PN-TPT unit cell, the supercell slab includes five layers of Cu(111) (245 atoms) and two TPT molecules (A and B) with a vacuum gap of 15 Å in the z-direction. Except for the bottom three layers of the Cu-slab, all atoms were allowed for geometry relaxation until the force tolerance on each atom was < 0.02 eV/Å. In the geometry optimization, only gamma Γ point was used to the Brillouin zone of the supercell, and significantly increased *k*-pint sampling 3×3×1 was used as the basis for the electronic and spin state calculations. Moreover, for the proper treatment of the electron correlation in the localized *d*-Co orbital, we used the Hubbard-like approximation in the DFT (DFT+U), U = 2.5 eV[38]. STM simulation was performed using the Tersoff-Hamann approximation[39], and the constant charge density contour was generated as the constant-current image in P4VASP[40]. In the simulated STM, the specific energies are selected to match experimental bias, and the positive and negative represent occupied and unoccupied states, respectively. The net charges calculation was by Bader Charge allocation[41].

### References

1. Barth, J. V. Fresh perspectives for surface coordination chemistry. *Surf Sci* **603**, 1533-1541 (2009).
2. Gutzler, R., Stepanow, S., Grumelli, D., Lingenfelder, M., Kern, K. Mimicking enzymatic active sites on surfaces for energy conversion chemistry. *Acc Chem Res* **48**, 2132-2139 (2015).
3. Gambardella, P. et al. Supramolecular control of the magnetic anisotropy in two-dimensional high-spin Fe arrays at a metal interface. *Nature materials* **8**, 189 (2009).

4. Umbach, T. R. et al. Ferromagnetic coupling of mononuclear Fe centers in a self-assembled metal-organic network on Au(111). *Phys Rev Lett* **109**, 267207 (2012).
5. Muller, K., Enache, M., Stohr, M. Confinement properties of 2D porous molecular networks on metal surfaces. *J Phys Condens Matter* **28**, 153003 (2016).
6. Khajetoorians, A. A., Wegner, D., Otte, A. F., Swart, I. Creating designer quantum states of matter atom-by-atom. *Nature Reviews Physics* **1**, 703-715 (2019).
7. Yan, L., Liljeroth, P. Engineered electronic states in atomically precise artificial lattices and graphene nanoribbons. *Advances in Physics: X* **4**, 1651672 (2019).
8. Barth, J. V. Molecular architectonic on metal surfaces. *Annu Rev Phys Chem* **58**, 375-407 (2007).
9. Grill, L., Dyer, M., Lafferentz, L., Persson, M., Peters, M. V., Hecht, S. J. N. n. Nano-architectures by covalent assembly of molecular building blocks. *Nature Nanotechnology* **2**, 687 (2007).
10. Dong, L., Gao, Z. A., Lin, N. Self-assembly of metal-organic coordination structures on surfaces. *Progress in Surface Science* **91**, 101-135 (2016).
11. Pawin, G., Wong, K. L., Kwon, K.-Y., Bartels, L. A homomolecular porous network at a Cu(111) surface. *Science* **313**, 961-962 (2006).
12. Piquero-Zulaica, I. et al. Precise engineering of quantum dot array coupling through their barrier widths. *Nature communications* **8**, 787 (2017).
13. Lin, T., Kuang, G., Shang, X. S., Liu, P. N., Lin, N. Self-assembly of metal–organic coordination networks using on-surface synthesized ligands. *Chemical Communications* **50**, 15327-15329 (2014).
14. Han, P., Weiss, P. S. Electronic substrate-mediated interactions. *Surface Science Reports* **67**, 19-81 (2012).
15. Crommie, M. F., Lutz, C. P., Eigler, D. M. Confinement of electrons to quantum corrals on a metal surface. *Science* **262**, 218-220 (1993).
16. Piquero-Zulaica, I. et al. Surface state tunable energy and mass renormalization from homothetic quantum dot arrays. *Nanoscale* **11**, 23132-23138 (2019).
17. Nowakowska, S. et al. Configuring Electronic States in an Atomically Precise Array of Quantum Boxes. *Small* **12**, 3757-3763 (2016).
18. Lyu, L. et al. Thermal-Driven Formation of 2D Nanoporous Networks on Metal Surfaces. *Journal of Physical Chemistry C* **123**, 26263-26271 (2019).
19. Lobo-Checa, J. et al. Band formation from coupled quantum dots formed by a nanoporous network on a copper surface. *Science* **325**, 300-303 (2009).
20. Shockley, W. On the surface states associated with a periodic potential. *Phys Rev* **56**, 317 (1939).
21. Kevan, S., Gaylord, R. High-resolution photoemission study of the electronic structure of the noble-metal (111) surfaces. *Physical Review B* **36**, 5809 (1987).
22. Hengsberger, M., Baumberger, F., Neff, H., Greber, T., Osterwalder, J. Photoemission momentum mapping and wave function analysis of surface and bulk states on flat Cu (111) and stepped Cu (443) surfaces: A two-photon photoemission study. *Physical Review B* **77**, 085425 (2008).
23. Scheybal, A. et al. Modification of the Cu(110) Shockley surface state by an adsorbed pentacene monolayer. *Physical Review B* **79**, 115406 (2009).
24. Piquero-Zulaica, I. et al. Effective determination of surface potential landscapes from metal-organic nanoporous network overlayers. *New Journal of Physics*, (2019).
25. Seufert, K. et al. Controlled Interaction of Surface Quantum-Well Electronic States. *Nano Letters* **13**, 6130-6135 (2013).
26. Ahsan, A. et al. Watching nanostructure growth: kinetically controlled diffusion and condensation of Xe in a surface metal organic network. *Nanoscale* **11**, 4895-4903 (2019).
27. Madueno, R., Räisänen, M. T., Silien, C., Buck, M. Functionalizing hydrogen-bonded surface networks with self-assembled monolayers. *Nature* **454**, 618 (2008).

28. Nowakowska, S. et al. Interplay of weak interactions in the atom-by-atom condensation of xenon within quantum boxes. *Nature communications* **6**, (2015).
29. Kawai, S. et al. Van der Waals interactions and the limits of isolated atom models at interfaces. *Nature communications* **7**, 1-7 (2016).
30. Cheng, Z. et al. Adsorbates in a box: titration of substrate electronic states. *Phys Rev Lett* **105**, 066104 (2010).
31. Pacchioni, G. E. et al. Two-Orbital Kondo Screening in a Self-Assembled Metal–Organic Complex. *Acs Nano* **11**, 2675-2681 (2017).
32. Nečas, D., Klapetek, P. Gwyddion: an open-source software for SPM data analysis. *Open Physics* **10**, 181-188 (2012).
33. de Abajo, F. G., Cordón, J., Corso, M., Schiller, F., Ortega, J. E. Lateral engineering of surface states–towards surface-state nanoelectronics. *Nanoscale* **2**, 717-721 (2010).
34. Kresse, G., Furthmüller, J. Efficiency of ab-initio total energy calculations for metals and semiconductors using a plane-wave basis set. *Computational materials science* **6**, 15-50 (1996).
35. Blöchl, P. E. Projector augmented-wave method. *Physical Review B* **50**, 17953-17979 (1994).
36. Perdew, J. P., Burke, K., Ernzerhof, M. Generalized gradient approximation made simple. *Phys Rev Lett* **77**, 3865 (1996).
37. Klimeš, J., Bowler, D. R., Michaelides, A. Van der Waals density functionals applied to solids. *Physical Review B* **83**, 195131 (2011).
38. Mann, G. W., Lee, K., Cococcioni, M., Smit, B., Neaton, J. B. First-principles Hubbard U approach for small molecule binding in metal-organic frameworks. *The Journal of chemical physics* **144**, 174104 (2016).
39. Tersoff, J., Hamann, D. R. Theory of the scanning tunneling microscope. *Physical Review B* **31**, 805 (1985).
40. P4VASP. http://www.p4vasp.at.
41. Sanville, E., Kenny, S. D., Smith, R., Henkelman, G. Improved grid‐based algorithm for Bader charge allocation. *Journal of computational chemistry* **28**, 899-908 (2007).
42. Jauernik, S., Hein, P., Gurgel, M., Falke, J., Bauer, M. Probing long-range structural order in SnPc/Ag(111) by umklapp process assisted low-energy angle-resolved photoelectron spectroscopy. *Physical Review B* **97**, 125413 (2018).

**Acknowledgments**

This work was supported by the Deutsche Forschungsgemeinschaft (DFG, German Research Foundation), TRR 173-268565370 Spin + X: spin in its collective environment (Project B05). B.S. acknowledges financial support by the Dynamics and Topology Center funded by the State of Rhineland Palatinate. Scientific Research Fund of Hunan Provincial Education Department of China (Grant No. 21B0548, Grant No. 19B159).

# Temperature-driven quantum confinements of surface electrons and adatoms in a weakly interacting 2D organic porous network

by
L. Lyu et al.

**Supplementary Information**

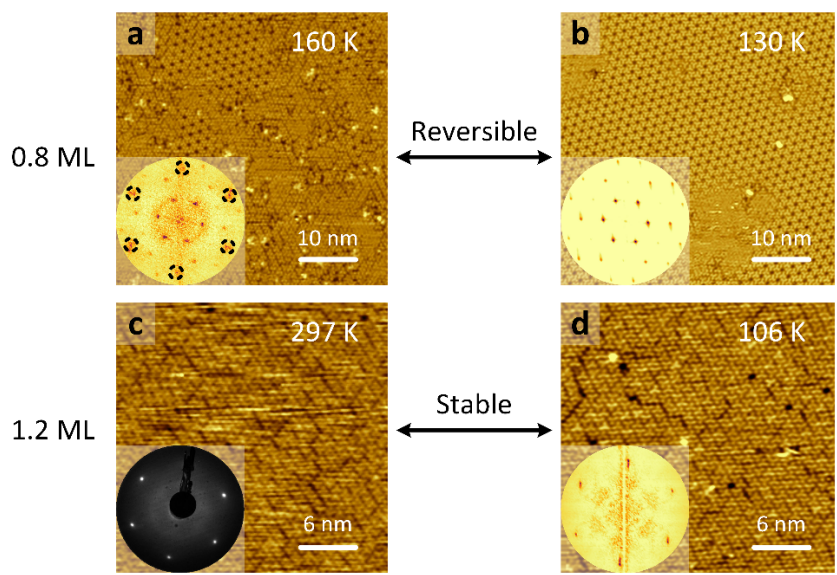

**Fig SI. 1 | Temperature and coverage dependent TPT phases on Cu(111). a,b,** At a coverage of 0.8 ML, the structures at the temperatures of (**a**) 160 K ($V_{tip}$ = + 0.16 V) and (**b**) 130 K ($V_{tip}$ = + 0.50 V). The lower-left insets are the corresponding FFT patterns (4 × 4 nm$^{-2}$). The black-circled spots in the FFT of **a** are from the close-packed TPT structure (CP-TPT), and the other spots correspond to the TPT porous network (PN-TPT). **c,d,** At a coverage of 1.2 ML, CP-phase remains in the temperature of (**c**) 297K and (**d**) 106 K. The spots in the LEED image (at 12 eV) of **c** is in agreement with the FFT spots (4 × 4 nm$^{-2}$) of **d**.

Deposition of 0.8 ML TPT on Cu(111), Fig. SI1a and SI1b show the TPT phases at 160 K and 130 K, respectively. At 160 K, a mixing structure includes the CP-TPT and PN-TPT, and the corresponding FFT also indicates two superimposed spots from the two phases. At 130 K, it shows a TPT porous structure in most of the area, but a transition appears in some fuzzy regions. Increasing to a higher coverage of 1.2 ML, a consistent CP-TPT is at 297 K (Fig. SI1c) and at 106 K (Fig. SI1d). It indicates that the additional molecules destroy the reversible phase transition.

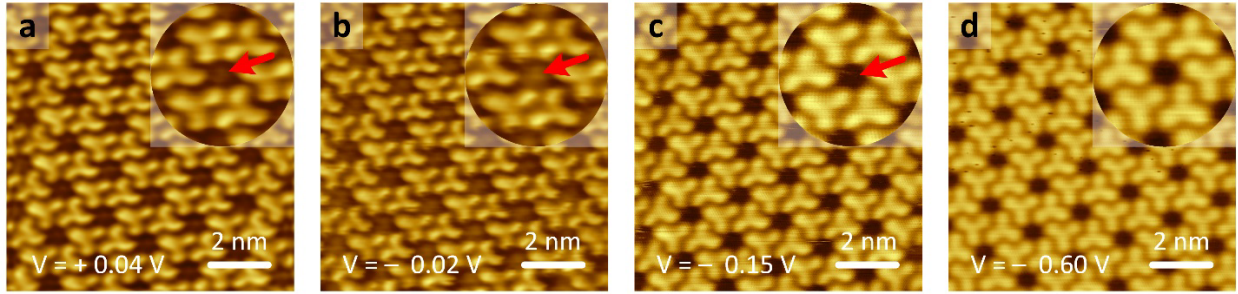

**Fig SI. 2 | STM tip bias ($V_{tip}$)-dependent SS electrons in TPT porous network. a-d** STM images (at constant-current mode) are acquired at the $V_{tip}$ of + 0.04 V, – 0.02 V, – 0.15 V and – 0.60 V, respectively. The upper-right inserts are the magnified porous units, in which the red arrows mark the domelike porous states. STM tunneling current ($I_t$), 0.07 nA in **a-d**.

The tip bias ($V_{tip}$)-dependent porous states are present in Fig. SI2a-d. An apparent domelike porous state occurs around the $V_{tip}$ = 0 V (Fig. SI2a and SI2b). That indicates a confined surface state (CSS) in each cavity. Due to a charge density contrast and a height difference between the porous state and TPT molecules, the CSS is almost invisible at the large $V_{tip}$ absolute values, as seen in Fig. SI2c and SI2d.

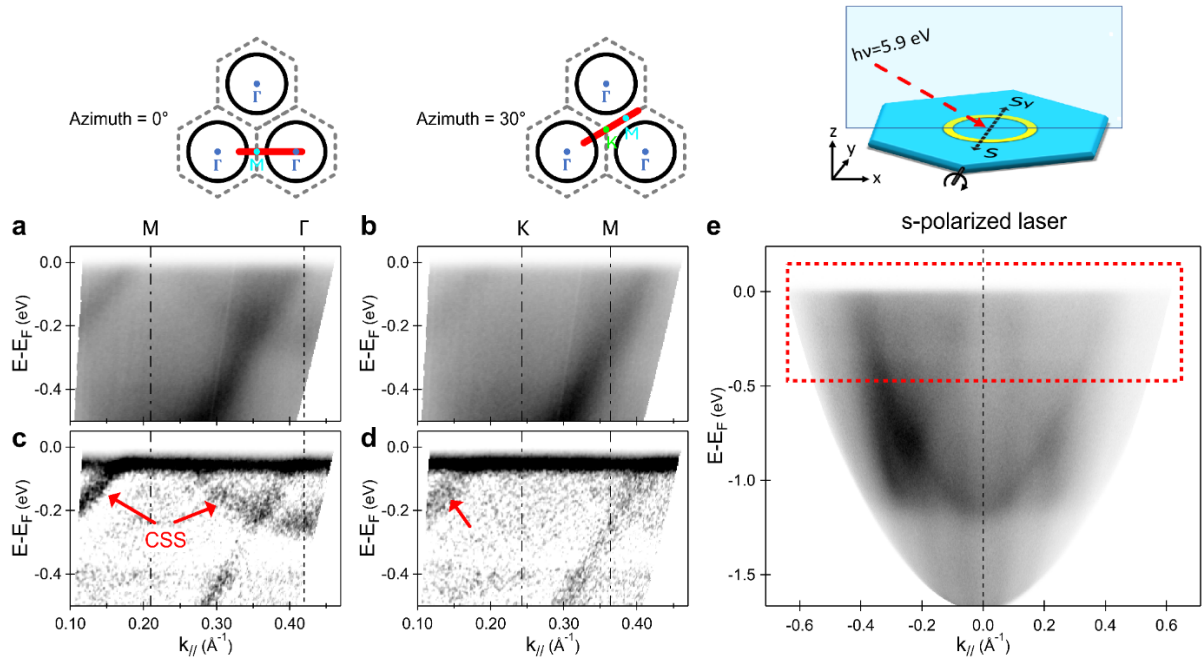

**Fig SI. 3 | ARPES bands of PN-TPT at different k$_{//}$ directions and laser polarizations. a,b,** Linearly *p*-polarized ARPES (hν = 5.9 eV) along with the $\overline{\Gamma M \Gamma}$ and $\overline{\Gamma K M}$ direction, respectively. The corresponding reciprocal direction is shown in the top schemas. **c,d,** The second derivative maps of **a** and **b**, improving the visualization of the SS replicas as marked by the red arrows. **e,** A linearly *s*-polarized ARPES (hν = 5.9 eV) along with the $\overline{\Gamma M \Gamma}$ direction, no SS signals occurs near the Fermi (red box region). A schematic of the *s*-polarized laser-ARPES setup is drawn on the top. All ARPES spectra were recorded at 30 K.

Due to the SS confinement in the periodic TPT porous network, the CSS replicas are located in each surface Brillouin zone of the porous momentum space. Along the $\overline{\Gamma M \Gamma}$ direction, the two CSSs appear in the adjacent BZs, as seen in Fig. SI3a and SI3c. Along the $\overline{\Gamma K M}$ direction, only one CSS is observed in the $\overline{\Gamma K}$ region (Fig. SI3b and SI3d), and the CSS in the $\overline{K M}$ region is above the Fermi[1]. Due to the transition matrix that describes the interaction between the light field and the material's electronic states[22], only *p*-polarized laser light ($p_z$ vector out-of-plane, Fig. 2a) can effectively excite SSs electrons. In contrast, there is no photoexcitation of the $p_z$-like SS for a *s*-polarized light (in-plane, top schematic in Fig. SI3e). Therefore, the CSSs in Fig. SI3e are invisible in the *s*-polarized laser ARPES measurement, in which the band structures arise from the back-folded Mahan cones[42].

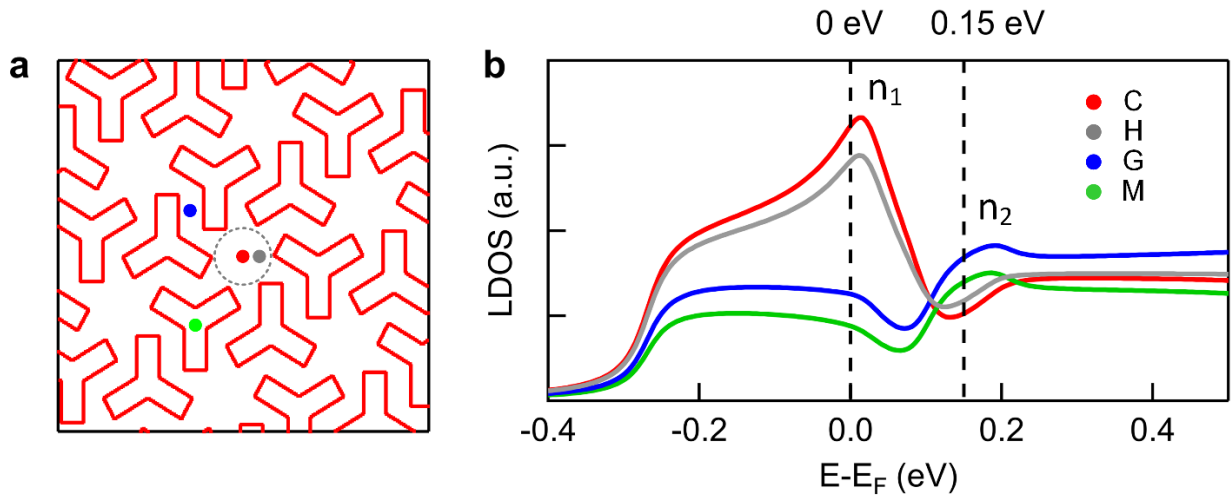

**Fig SI. 4 | EPWE simulation for the LDOS spectra of SS in a PN-TPT model. a,** PN-TPT model for the EPWE simulation. **b,** The EPWE simulated LDOS spectra of SS at the four positions (marked in **a**): porous center (C, red spot), porous halfway (H, grey spot), intermolecular gap (G, blue spot) and top of the molecule (M, green spot). The two peaks (marked by dash lines) at the $E - E_F$ = 0 eV and $E - E_F$ = + 0.15 eV are located in the discrete CSS eigenstates ($n_1$ and $n_2$)

Fig. SI4a is a simplified model for the PN-TPT structure. Fig. SI4b shows the EPWE simulated LDOS spectra for the confined SS. Comparing the LDOS spectra at the four positions, the $n_1$ state

is located in the cavity (the dash-circle region in Fig. SI4a), and the n$_2$ state is mainly in the intermolecular gaps. They correspond to the LDOS maps in Fig. 2i and 2j.

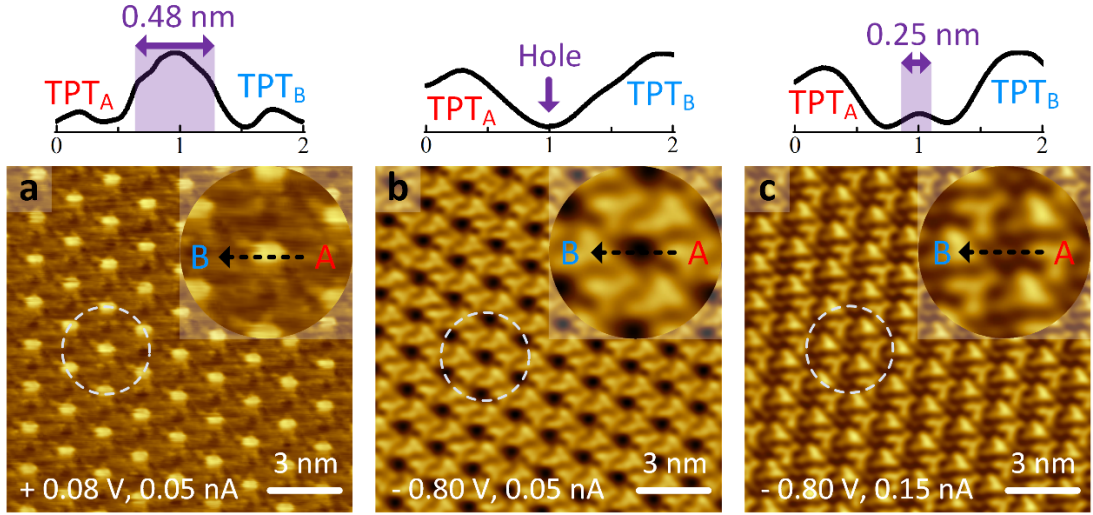

**Fig SI. 5 | STM topographies of the full-doped Co in PN-TPT at different V$_{tip}$ and I$_t$. a-c,** The (V$_{tip}$, I$_t$) is (+ 0.08 V, 0.05 nA), (− 0.80 V, 0.05 nA) and (− 0.80 V, 0.15 nA), respectively. The upper-right inserts are the magnified porous unit in the white circle, showing the (V$_{tip}$, I$_t$)-dependent porous structure. The top line profiles cross the cavity, along the black arrow from TPT$_A$ to TPT$_B$. All STM images at 106 K.

Fig. SI5 shows the full-doped Co atoms in PN-TPT at a series of V$_{tip}$ and I$_t$. For a constant current STM, the $I_t$ is proportional to the $V_{tip} \cdot e^{-2\kappa d}$, where the $k$ is constant, and $d$ is the tip-sample distance. In Fig. SI5a, the STM image at V$_{tip}$ = 0.08 V shows a weak TPT state but a protruding signature of the porous Co cluster, and the line profile shows the cluster diameter of 0.48 nm. In Fig. SI5b, the increased V$_{tip}$ (at − 0.80 V) results in a stronger TPT state but an empty porous structure. In Fig. SI5c, the V$_{tip}$ remains the same at − 0.80 V, and an increase of the I$_t$ (from 0.05 nA to 0.15 nA) causes the tip to approach the surface (smaller $d$), resulting in the reappearance of the Co cluster in the cavity (size of 0.25 nm in the line profile). The process of the Co cluster's disappearance and reappearance indicates that the adsorption height of the Co atoms is lower than that of the TPT layer (~ 3 Å).

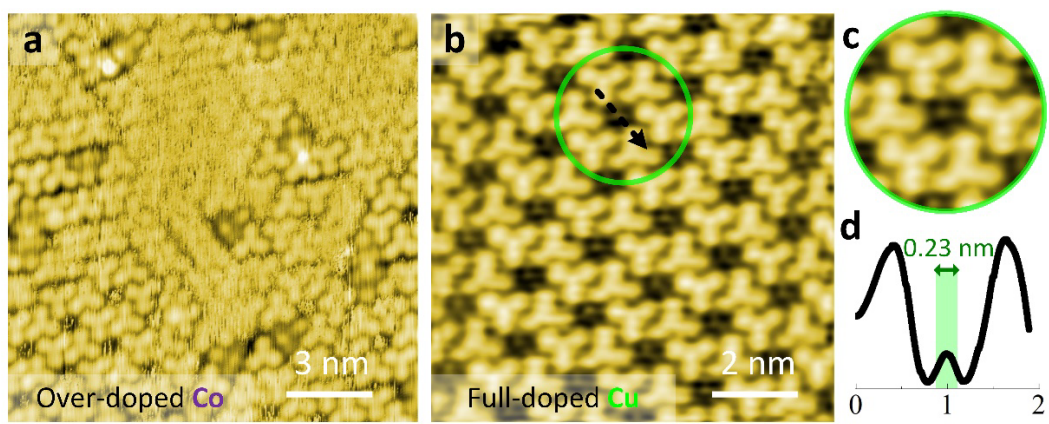

**Fig SI. 6 | Confinement of Cu adatoms in TPT porous network. a**, STM topograph ($V_{tip}$ = + 0.20 V) of an over-doped Co atom (0.08 ML) in PN-TPT. **b**, STM topograph ($V_{tip}$ = + 0.30 V) of full-occupied Cu atoms in PN-TPT. **c**, The magnified occupancy unit in the green circle. **d**, The line profile crossing the porous Cu cluster (black dash arrow) shows the Cu cluster with a size of 0.23 nm. All STM images were measured at 106 K.

Fig. SI6a shows a destructed PN-TPT structure (named over-doped Co) by the excess Co deposition (0.08 ML). In Fig. SI6b-d, full-doped Cu atoms appear in the PN-TPT. Keeping the same $V_{tip}$ and $I_t$ as Fig. 3c, a small Cu cluster spot is observed in each cavity. However, the line profile shows the confined Cu cluster at a size of 0.23 nm, much smaller than the full-doped Co cluster (0.32 nm), and the surrounding TPT molecules keep an identical state. This indicates a weaker interaction between the porous Cu cluster and the surrounding TPT molecules.

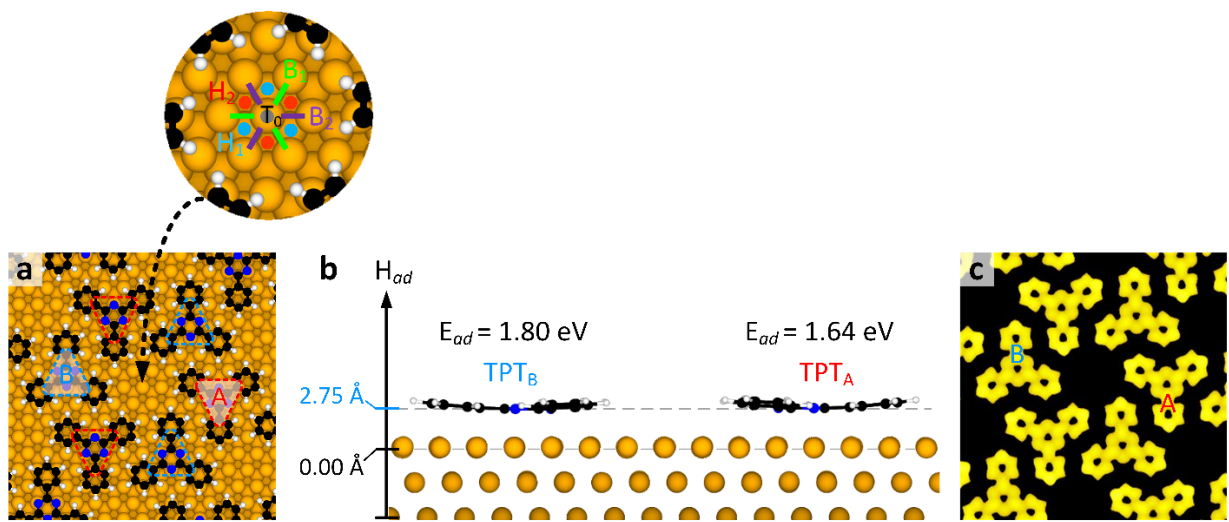

**Fig SI. 7 | Structural simulation of TPT porous network on Cu(111).** Top view (**a**) and side view (**b**) of the optimized PN-TPT on Cu(111). Two TPT types (TPT$_A$ and TPT$_B$) on different surface stacks: the TPT triazine center on the FCC site (A, marked by the blue triangles) and on the HCP site (B, red triangles). The top magnified unit

shows the symmetry locations in a cavity: center ($T_0$, grey spot), bridge site 1 ($B_1$, three green rods), bridge site 2 ($B_2$, three purple rods), hollow site 1 ($H_1$, three blue spots) and hollow site 2 ($H_2$, red spots). The adsorption heights ($H_{ad}$) and adsorption energies ($E_{ad}$) of $TPT_A$ and $TPT_B$ are labelled in **b**. **c**, A simulated STM image for **a** structure, calculated at a constant current mode with the energy of 0.6 eV below the Fermi. The scales of **a,c** are 3.6 × 3.6 nm$^2$.

In Fig. SI7a, b, DFT simulation shows the optimized structure of PN-TPT on the Cu(111) surface. The two types of TPT molecules ($TPT_A$ and $TPT_B$) adsorb on the different surface locations. In Fig. SI7b, $TPT_A$ and $TPT_B$ are almost the same height ($H_{ad}$), but the adsorption energy ($E_{ad}$) of $TPT_A$ is smaller than $TPT_B$. The simulated STM image (Fig. 5c) shows an identical electron density for both types of molecules, which is in agreement with the STM in Fig. 1c. As a result of the two types of TPT surrounding a cavity, the structural symmetry of a cavity degenerates from a $C_{6v}$ to a $C_{3v}$, and the highly symmetric adsorption sites in a cavity can be identified as $T_0$, $B_1$, $B_2$, $H_1$ and $H_2$ (colour marks in the magnified unit). The adsorbed Co atoms are most likely to be located at these sites.

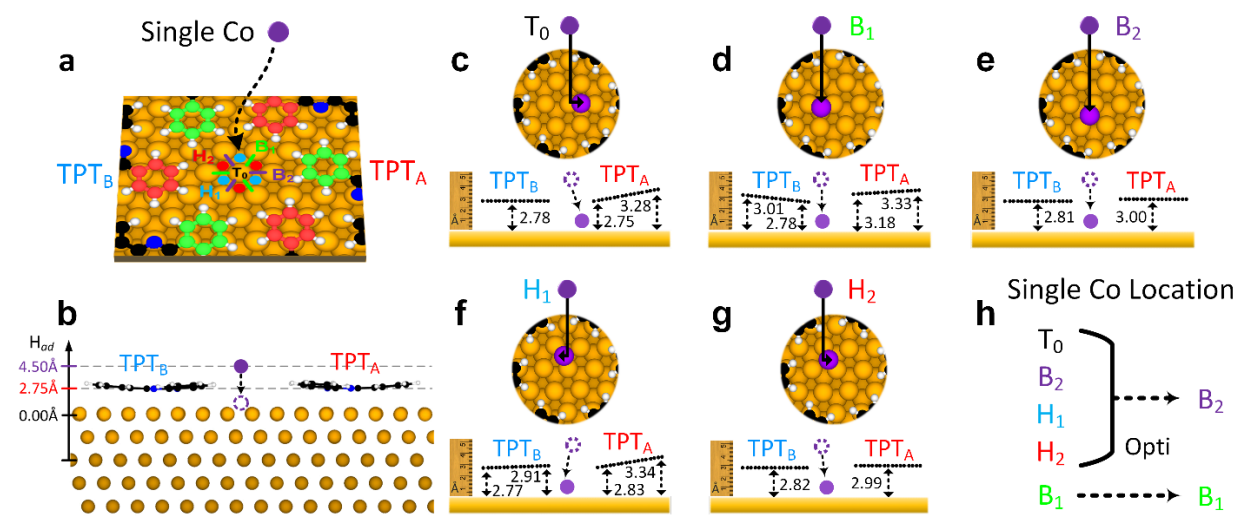

**Fig SI. 8 | Structural simulation for single Co atom into PN-TPT. a,b**, The schematics of a single Co atom into the optimized PN-TPT (top and side views). The initial Co atom is set to 4.5 Å on top of the five sites: $T_0$, $B_1$, $B_2$, $H_1$ and $H_2$. **c-g**, The optimized locations of the Co into PN-TPT. The upper parts are the models of the adsorbed Co position, and the lower parts show the tilts of $TPT_A$ and $TPT_B$ after the Co doping. **h**, The single Co atom location in different structures.

DFT simulation of a single Co atom trapped in the PN-TPT. As shown in Fig. SI8a, b, the Co atom is firstly placed out of the cavity, and the position is on the top of the five symmetry sites, i.e., $T_0$, $B_1$, $B_2$, $H_1$ and $H_2$. After the structural simulations (Fig. SI8c-g), the Co is optimized to the preferred location, and the doping process induces the different molecular tilts of $TPT_A$ and $TPT_B$. In Fig. SI8h, the Co atom placed on the top of $T_0$, $B_2$, $H_1$ and $H_2$, is preferentially located in the porous $B_2$ site. On the TPT side, the adsorption on the $T_0$, $B_1$ and $H_1$ sites leads to the apparent tilts in $TPT_A$ or $TPT_B$. However, considering the under-doped case in STM (Fig. 3b), the surrounding TPT molecules do not show any obvious difference. This case is consistent with the $B_2$ and $H_2$ doping, where the Co is preferentially located at the $B_2$ site, and $TPT_A$ and $TPT_B$ are flat structures with a slight height difference ($\Delta H_{ad} < 0.20$ Å).

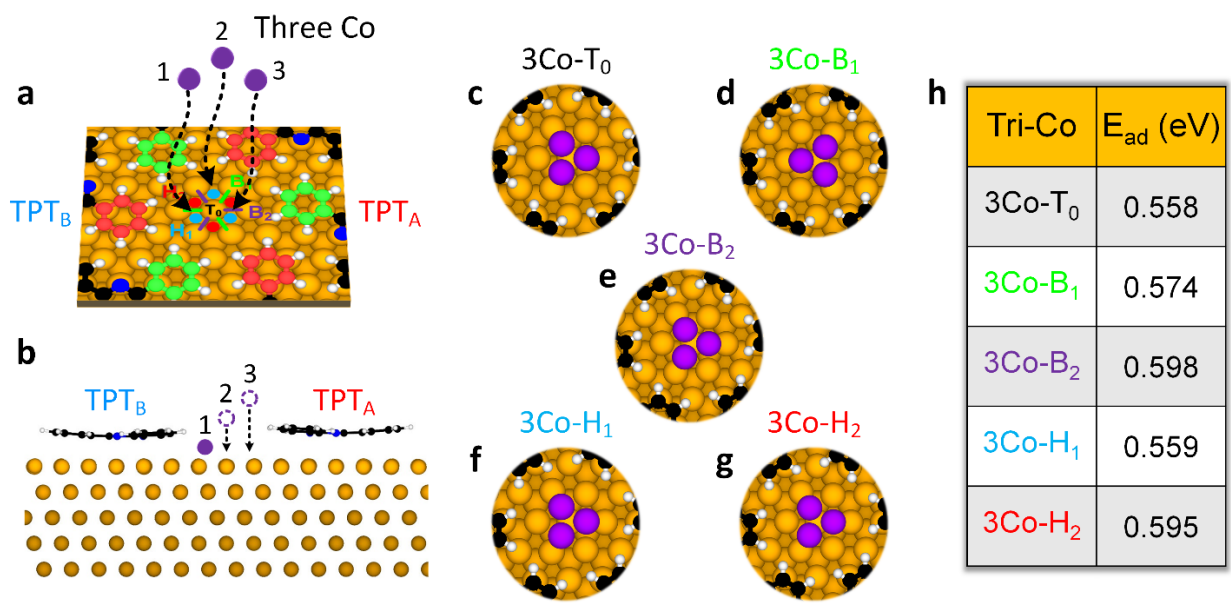

**Fig SI. 9 | Structural simulation for three Co atoms into PN-TPT. a,b**, The schematics of three Co atoms into the optimized PN-TPT (top and side views). The initial positions of three Co are sequentially placed on the top of the equivalent sites in $T_0$ (3Co-$T_0$), $B_1$ (3Co-$B_1$), $B_2$ (3Co-$B_2$), $H_1$ (3Co-$H_1$) and $H_2$ (3Co-$H_2$). **c-g**, The optimized Co locations of the 3Co-$T_0$, 3Co-$B_1$, 3Co-$B_2$, 3Co-$H_1$ and 3Co-$H_2$. **h**, The adsorption energies ($E_{ad}$) for the **c-g** structures, $E_{ad} = \frac{1}{3}(3E_{Co} + E_{PN-TPT} - E_{3Co-PNTPT})$.

In Fig. SI9, the simulation is shown for the full-doped Co in PN-TPT. Fig. SI9a and 9b show the trapping process of three Co atoms. The three Co atoms are sequentially placed on the top ($H_{ad}$ = 4.5 Å) of the $T_0$, $B_1$, $B_2$, $H_1$ and $H_2$ sites. For example, $H_1$ has three symmetry sites (blue dots) in a cavity. Three Co atoms, placed on the top of each $H_1$ site, are optimized into the cavity one by one. The optimized porous structures are 3Co-$T_0$, 3Co-$B_1$, 3Co-$B_2$, 3Co-$H_1$ and 3Co-$H_2$ in Fig. SI9c-g, respectively. In 3Co-$T_0$, 3Co-$B_2$, 3Co-$H_1$ and 3Co-$H_2$, the Co atoms favour the three $B_2$ sites. Based on the adsorption energies ($E_{ad}$) in Fig. SI9h, 3Co-$B_2$ and 3Co-$H_2$ have a bigger $E_{ad}$ than others. 3Co-$H_2$ and 3Co-$B_2$ are almost the same geometry after optimization. Therefore, the 3Co-$B_2$ is proposed as the optimum structure. Moreover, adding another Co in the 3Co-$B_2$ (four

Co doping) is also calculated, and the optimized structure (not shown) reveals that the fourth Co atom is located on the top of the first three atoms. As a result, a bilayer Co cluster occurs in the cavity for the four Co atoms system. However, the height of the bilayer Co is higher than that of the TPT layer, which is inconformity to the experimental results in Fig. SI5. Therefore, the structure of full-doped Co in PN-TPT appoints to the three Co-doped 3Co-B$_2$.

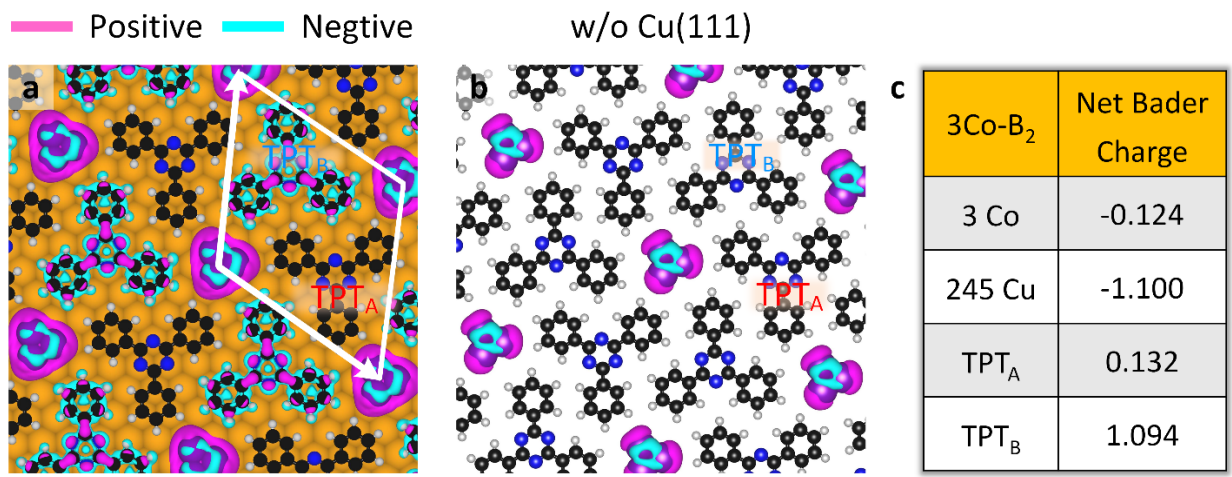

**Fig SI. 10 | Charge transfer simulation of full-doped Co in TPT porous network. a**, Charge density differences (CDD) in full-doped Co structure, i.e., 3Co-B$_2$. Positive and negative values correspond to accumulated and depleted charges, and the white rhombus represents a porous unit cell. **b**, The CDD of removing the Cu substrate from **a** (w/o Cu(111)). **c**, The calculation of net Bader charges in one unit cell, components including 3 Co atoms, 245 Cu atoms, 1 TPT$_A$ and 1 TPT$_B$. Isosurface values in **a,b**, +/– 0.0015 to positive and negative charges, respectively. The scales of **a,b**, 3.6 × 3.6 nm$^2$.

Fig. SI10a calculates a charge density difference for the full-doped 3Co-B$_2$ system. The net charges accumulate in the Co clusters and the TPT$_B$ molecules. When removing the Cu substrate from the 3Co-B$_2$ (Fig. SI10b), the charges are only localized in the Co clusters. It indicates the charge transfer between the Co cluster and TPT molecules is through the Cu substrate. From the net Bader charges in Fig. SI10b, electrons from the doped Co cluster and substrate are mainly injected into the TPT$_B$ molecule. The strength of the charge accumulation on the molecules depends on the molecule-surface coupling. Hence, the uplifted TPT$_A$ molecules have negligible charge states.

**Supplementary References**

[1]. Seufert, K. et al. Controlled Interaction of Surface Quantum-Well Electronic States. *Nano Letters* **13**, 6130-6135 (2013).